\documentclass{llncs}
\usepackage{makeidx}  % allows for indexgeneration
\usepackage{graphicx}
\usepackage[caption=false,font=footnotesize]{subfig}
\usepackage{tabularx}
\newcolumntype{Y}{>{\centering\arraybackslash}X}
\usepackage{amssymb}
\usepackage{amsmath}
\usepackage{amssymb}
\usepackage{amsmath}
\usepackage{xspace}
\usepackage{cite}
\usepackage{color}
\usepackage{url}
\usepackage{epsfig}
\usepackage{psfrag}       %latex text in .eps files
\usepackage[ruled,noend,linesnumbered]{algorithm2e}%noend
\usepackage{listings}
\usepackage{ulem} 
\usepackage{picins}

\definecolor{light-gray}{gray}{0.5}

\newcommand{\DN}{{Dynamic Network}\xspace}
\newcommand{\DNs}{{Dynamic Networks}\xspace}
\newcommand{\ADN}{{Anonymous Dynamic Network}\xspace}
\newcommand{\ADNs}{{Anonymous Dynamic Networks}\xspace}
\newcommand{\IC}{\textsc{Incremental Counting}\xspace}
\newcommand{\footnotenonumber}[1]{{\def\thempfn{}\footnotetext{#1}}}
\newcommand{\mig}[1]{\textcolor{blue}{#1}}
\newcommand{\miguel}[1]{\textcolor{blue}{#1}}
\renewcommand{\mig}[1]{#1}
\renewcommand{\miguel}[1]{#1}
\begin{document}
\mainmatter              % start of the contributions
\title{
Counting in Practical \ADNs is Polynomial
}
\titlerunning{Counting in Practical \ADNs is Polynomial}  % abbreviated title (for running head)
\author{
Maitri Chakraborty\inst{1}
\and
Alessia Milani\inst{2}
\and
Miguel A. Mosteiro\inst{1}
}
\authorrunning{M. Chakraborty, A. Milani, and M. A. Mosteiro} % abbreviated author list (for running head)
%\tocauthor{Ivar Ekeland, Roger Temam, Jeffrey Dean, David Grove, Craig Chambers, Kim B. Bruce, and Elisa Bertino}
%
\institute{
Kean University,
Union, NJ, USA,\\
\email{\{chakrabm,mmosteir\}@kean.edu}
\and
LABRI, University of Bordeaux, INP,
Talence, France,\\
\email{milani@labri.fr}
}

\maketitle              % typeset the title of the contribution

\begin{abstract}
\ADNs is a harsh computational environment due to changing topology and lack of identifiers. Computing the size of the network, a problem known as Counting, is particularly challenging because messages received cannot be tagged to a specific sender. Previous works on Counting in \ADNs do not provide enough guarantees to be used in practice. Indeed, they either compute only an upper bound on the network size that may be as bad as exponential, or guarantee only double-exponential running time, or do not terminate, or guarantee only eventual termination without running-time guarantees. Faster experimental protocols do not guarantee the correct count. 

Recently, we presented the first Counting protocol that computes the exact count with exponential running-time guarantees. \miguel{The protocol requires the presence of one leader node and knowledge of any upper bound $\Delta$ on the maximum number of neighbors that any node will ever have.}
In the present work, we complement the latter theoretical study evaluating the performance of such protocol in practice. We tested a variety of network topologies that may appear in practice, including extremal cases such as trees, paths, and continuously changing topologies. We also tested networks that temporarily are not connected. Our simulations showed that the protocol is polynomial for all the inputs tested, paving the way to use it in practical applications where topology changes are predictable. The simulations also provided insight on the impact of topology changes on information dissemination.

To the best of our knowledge, this is the first experimental study that shows the possibility of computing the exact count in polynomial time in a variety of \ADNs that are worse than expected in practice.

\end{abstract}

\footnotenonumber{We thank David Joiner for assisting us in using the Kean Terascale Cluster (KTC) for our simulations.}

%
%!TEX root = main_llncs.tex

\section{Introduction}
\label{sec:intro}

Recently, a restrictive \ADN model where node identifiers are not available and topology changes frequently has attracted a lot of attention. 
With respect to topology changes, the \ADN model is well motivated by mobility and unreliable communication environments. With respect to node identifiers, although they are usually available in present networks (or labels\footnote{Throughout, we use ``identifiers'' or ``labels'' indistinctively.} are defined at startup), in future massive networks it may be necessary or at least convenient to avoid them to facilitate mass production.
%Moreover, even if identifiers are still available in future networks, they may not be unique or it may be convenient to ignore them under highly dynamic conditions. 

In particular, the seemingly simple problem of \emph{Counting} the number of nodes is challenging in \ADNs. Indeed, when two nodes communicate, it is not known whether they have communicated previously or not, which makes difficult to count. However, Counting is a fundamental problem in distributed computing because the network size is used to decide termination of protocols. 

The literature on Counting in \ADNs (cf.~\cite{conscious,oracle,spirakis,experimentalConscious,opodisCounting}) focuses on distributed protocols for broadcast networks in slotted-time scenarios, assuming adversarially that topology may change completely all the time. Fruitful results obtained so far showed that Counting is feasible in \ADNs, paving the way to understand the cost of anonymity in \DNs, but it is still not known whether those protocols are practical.
Indeed, the protocol in~\cite{spirakis} computes only an upper bound on the network size that may be as bad as exponential. The protocols in~\cite{conscious} compute the exact count, but one guarantees only double-exponential running time and the other, called \emph{unconscious}, does not terminate. Another protocol~\cite{oracle} is shown to have eventual termination, but without running-time guarantees.

%On the practical side, experimental protocols presented are heuristic and do not guarantee the correct count~\cite{experimentalConscious}.
%In brief, nodes estimate the maximum number of neighboring nodes keeping track of the maximum seen so far. Unfortunately, in the worst case this approach could yield an arbitrarily wrong estimate, yielding a wrong overall count \mig{(\emph{``... however, the error rate becomes high when we consider sparse and extremely disconnected graph instances or regular topologies.''}~\cite{experimentalConscious})}. In this work, we focus in protocols that guarantee the correct computation.

On the practical side, in~\cite{experimentalConscious} the \emph{unconscious} protocol~\cite{conscious} is augmented with a termination heuristic that allows the leader to decide when to stop. They show experimentally that the algorithm converges to a decision in a number of rounds that is linear in the size of the system. Unfortunately, it ensures a correct count only on dense graphs. In the worst case this approach could yield an arbitrarily wrong estimate, yielding a wrong overall count (\emph{``... however, the error rate becomes high when we consider sparse and extremely disconnected graph instances or regular topologies.''}~\cite{experimentalConscious}). 

%In this work, we focus in protocols that guarantee the correct computation.
%\ale{In this work, we present a protocol that computes in polinomial time the exact count for a variety of network topologies that may appear in practice. It is derived from the protocol, called \emph{\IC},  we recently presented in~\cite{opodisCounting}. }

Recently, we presented the first Counting protocol that computes the exact count with exponential running-time guarantees in~\cite{opodisCounting}. The protocol, called \emph{\IC}, achieves a speedup over its predecessors by trying candidate sizes incrementally. 
%Thus, we name it here \emph{\IC} for easy reference. 
The analysis of \IC in~\cite{opodisCounting} also exposed the bottleneck for further speedup. Indeed, \IC and previous protocols include a \emph{collection} phase where nodes disseminate some value in a gossip-based fashion. Under adversarial topology changes, the best theoretical upper bound known for such collection is exponential, whereas the only lower bound known is the trivial lower bound for dissemination, i.e. the dynamic diameter. 
Moreover, even restricting to gossip-based protocols, it is not known if there exist adversarial topologies such that collection requires exponential time. 
Thus, whether the protocol performs better than exponential in practice is an important question.
%of paramount importance. 

In the present work, we complement the theoretical study in~\cite{opodisCounting} evaluating thoroughly the performance of \IC in practice. 
We tested a variety of network topologies that may appear in practice, including random, best case, and worst case. In particular, we tested (1) tree topologies carefully drawn uniformly at random from the equivalence classes defined by isomorphisms, (2) star topologies, and (3) path topologies. The simulations parameters include the size of the network $n$, an upper bound on the number of neighboring nodes $\Delta$, and the period of time without topology changes $T$, including the extremal cases when topology changes continuously and a static topology. We also tested (4) dynamic networks with random graph topologies that may be disconnected, which are relevant given that previous works~\cite{experimentalConscious} do not guarantee the correct answer under disconnection.

It should be noticed that, for the purpose of dissemination in this model, tree inputs are intuitively worse than graphs, the worst case being a path. In that sense, our experimental evaluation is more challenging for the algorithm than previous works~\cite{experimentalConscious} where the inputs considered were various versions of random graphs that are unlikely to be trees. 

For all the topologies and parameter combinations evaluated, \IC has proven to be polynomial in our simulations. These results motivate the application of \IC to practical \ADNs. 
%Indeed, the running times obtained statistically in the centralized simulator may be used to synchronize each phase of the protocol in a distributed implementation. 
Indeed, our simulations show that, for the inputs tested, $\Delta n^4$ is a loose upper bound on the running time of \IC. Hence, for applications where the input behavior is similar (and again we emphasize that we have used inputs that are bad for \IC), the collection phase of \IC may be stopped after $\Delta k^4$ iterations, where $k$ is the size estimate, to obtain a protocol that can be used in practice (refer to~\cite{opodisCounting} for details).  
%\sout{\miguel{Moreover, the results obtained provide enough intuition to conjecture that \IC is polynomial even in the worst case, although a formal proof of such conjecture is open.}}

Our simulations also provide insight on the impact of network dynamics in the dissemination of information by gossip-based protocols. Indeed, our results showed that, on average, network changes speed up the computation, as long as those changes are uniform throughout the network. That is, highly dynamic topologies help rather than being a challenge as in worst-case theoretical analyses.

The rest of the paper is organized as follows. 
After formally defining the model and the problem in Section~\ref{sec:model}, we detail the changes introduced into \IC to run our simulations in Section~\ref{sec:protocols}.
The input topologies tested and the simulation platform used are presented in Sections~\ref{sec:input} and~\ref{sec:platform}.
Finally, we discuss the results obtained and our conclusions in Section~\ref{sec:discussion}. 
Beyond the overview of previous work on Counting included above and the references therein, other related work may be found in a survey on Dynamic Networks and Time-varying Graphs by Casteigts et al.~\cite{arnaudSurvey}. 

%!TEX root = main_llncs.tex

\section{The \ADN Model and the Counting Problem}
\label{sec:model}

The \ADN model is defined as follows. We consider a network composed by a set $V$ of $n$ nodes. For reference, we define node labels $\{1,2,\dots,n\}$. However, nodes do not have identifiers or labels that may be used in the computation. That is, nodes cannot be distinguished, except for one node $\ell$ that is called the \emph{leader}. As shown in~\cite{spirakis}, Counting is not solvable in Anonymous Networks without the presence of a leader, even if the topology does not change.
If a given pair of nodes $i,j\in V$ is able to communicate directly, we say that there is a \emph{link} among them, and we say that $i$ and $j$ are \emph{neighbors}. 

The communication proceeds in synchronous \emph{rounds} through broadcast in symmetric links. That is, at each round, a node $i$ broadcasts a message to its neighbors and simultaneously receives the messages broadcast in the same round by all its neighbors. Then, each node makes some local computation (if any).

As customary in the \ADNs literature, we assume that the time taken by computations is negligible with respect to communication. Thus, to evaluate performance, we count the number of communication rounds to complete the computation.

The set of links among nodes at round $r$ is denoted as $E(r)$. For any given node $i$, we denote the set of neighbors of $i$ at round $r$ as $N(i,E(r))$. If the particular round is clear from context we will refer to the set of neighbors simply as $N(i,E)$ or $N(i)$ indistinctively.  
%It was conjectured in~\cite{spirakis} that any non-trivial computation is impossible without knowledge of some network characteristics. Such conjecture has not been proved or disproved. Thus,  as in previous work~\cite{conscious,opodisCounting},
We assume that there is an upper bound on the size of the neighborhood of any node that is known to all nodes. That is, there is a value $\Delta\leq n-1$ that may be used by the protocol such that, for any round $r$ and any node $i\in V$, it is $N(i,E(r))\leq \Delta$. Indeed, $\Delta$ is used in \IC as described in Section~\ref{sec:protocols}. Other than the upper bound $\Delta$, nodes do not have any other information of the network topology and/or dynamics. Thus, \IC does not use any other information.

Previous work~\cite{conscious,opodisCounting} also assume knowledge of this upper bound on the neighborhood. This is because it was conjectured in~\cite{spirakis} that any non-trivial computation is impossible without knowledge of some network characteristics. This conjecture has been recently disproved \cite{LunaB15}. On the other hand, the algorithm used to prove the result has exponential 
time and space complexity.

The set of links is dynamic. That is, they may change from one round to another. In~\cite{opodisCounting,conscious,spirakis}, Counting protocols have been analyzed assuming that in each round a new set of links may be chosen adversarially, as long as the network is connected.% ($\Delta\geq 2$). 
This model was presented in~\cite{KuhnLO2010} as $1$-interval connectivity model. 
Our simulations showed that if a new set of links is chosen uniformly at random for each round, the dissemination of information towards the leader is indeed faster than if changes are less frequent. Hence, for our simulations we generalize the connectivity model assumed in~\cite{opodisCounting,conscious,spirakis} as follows. We say that the network is \emph{$T$-stable} if, after a topology change, the set of links does not change for at least $T$ rounds. More formally, for any pair of rounds $r_i,r_j$ such that $0<r_i<r_j$, $E(r_i)\neq E(r_j)$, and for any $r_i\leq r_k<r_j$ it is $E(r_i)=E(r_k)$, then $r_j-r_i\geq T$. 
\mig{
In contrast, in $T$-interval connected networks it is assumed that for \emph{any} sequence of $T$ rounds, there is a stable %underlying graph 
\miguel{set of links}
spanning all nodes.
}

Notice that for connected networks both models are the same for $T=1$,
%. That is, a $1$-stable network is $1$-interval connected.
but for $T>1$ on tree topologies (most of our inputs), $T$-stable networks restrict less the adversary than $T$-interval connectivity networks. 
Indeed, if the topology is always a tree, which is a worst case scenario for dissemination, $T$-interval connectivity enforces a static network, whereas $T$-stability allows to change the tree every $T$ rounds. 
%Another way to notice why $T$-stability is less restrictive on trees is to notice that, out of all the possible blocks of $T$ rounds starting at each round, $T$-stability does not allow changes only in a $1/T$ fraction, whereas $T$-interval connectivity enforces a stable tree for all blocks of $T$ rounds. 
In this work, we study $T$-stable networks and we evaluate a range of values for $T$, from $T=1$ up to a static network. 
%\ale{We assume that the specific value of $T$ is not known beforehand.}  %So, we consider all cases.
%Moreover, in the general version of $1$-interval connectivity, which requires that for any $T$ consecutive rounds there is a stable underlying spanning graph,

%In $T$-stable networks nodes may use the message count after communication in one round to make decisions for the next round if the set of links will remain the same. However, we assume that the specific value of $T$ is not known beforehand.That is, the Counting protocol cannot use this information because in the worst case the topology will change. Within the same round, after communication, the protocol may use the number of messages received in the computation, but the number of neighbors for the next round is unknown.

We complete the section with the definition of the problem in~\cite{opodisCounting}: ``An algorithm is said to solve the \emph{Counting} problem if whenever it is executed in a \DN comprising $n$ nodes, all nodes eventually terminate and output $n$.''

%!TEX root = main_llncs.tex
\section{\IC Protocol Simulator}
\label{sec:protocols}

\IC~\cite{opodisCounting} is a distributed protocol that relies on the presence of a leader node. 
%The assumption is justified by the impossibility result proved in~\cite{spirakis}, where it was shown that the Counting problem is not solvable in Anonymous Networks without a \emph{leader}, even if the topology does not change. 
%The \IC algorithms for the leader (Algorithm~\ref{algo:leader}) and non-leader nodes (Algorithm~\ref{algo:no-leader}) are included in the Appendix for self-containment. 
Thus, the protocol includes algorithms for the leader and non-leader nodes. 
Both algorithms are composed by a sequence of synchronous iterations. In each iteration the candidate size is incremented and checked to decide whether it is correct or not. 

Each of the iterations in \IC is divided in three phases: collection, verification, and notification. In the collection phase, nodes are initially assigned a unit value, called energy. Then, iteratively nodes disseminate a fraction of their energy towards the leader. The collection phase terminates when the leader has collected enough energy to know that if the current guess of the system size is correct, then there is no node in the system with residual energy greater than a given threshold. In the verification phase, nodes inform the leader whether some node has energy above the threshold, which would mean that the candidate size is wrong. Should the candidate size be correct, in the last phase all nodes are notified that the computation is complete.   
Each of these phases is composed by a fixed number of communication rounds so that the synchronization of the distributed computation is given. 
The number of rounds of each phase is a function of the candidate size. 

\miguel{\IC runs for a fixed number of rounds for each phase. Given that the upper bound on the number of rounds needed for each phase proved in~\cite{opodisCounting} is exponential, a simulation of \IC as in~\cite{opodisCounting} would yield exponential time. The purpose of our simulations in the present work is to evaluate whether such upper bound is loose in practice.} 
So, 
%To evaluate if \IC is faster than exponential in practice, 
rather than running each phase for a fixed number of communication rounds, we do it until a condition suited for each phase is violated (cf. Algorithm~\ref{simprotocol}), and we count the number of communication rounds to complete the computation.
\miguel{Consequently, our simulator is necessarily centralized to check such condition, but again, these changes are introduced to obtain experimentally a tighter upper bound on practical inputs, rather than to provide a practical implementation of \IC. As explained before, a practical distributed \IC protocol must be implemented as in~\cite{opodisCounting}, that is, executing each phase for a fixed number of rounds, but our simulations provide a polynomial bound on that number.}

%Running simulations of \IC as is to evaluate time performance would not provide any useful information. First, because the number of communication rounds of all three phases is fixed, and consequently the time taken would be just the time to execute them. And second, because the best theoretical upper bound known for the number of collection rounds needed is exponential, and we aim to evaluate whether in practice the protocol could be run in polynomial time. 
%%MM:removed because it de-motivates the study of Gnp 
%%And third, because our evaluation inputs include topologies that may be temporarily disconnected, and given that the period of disconnection is bounded only stochastically, a protocol that runs for a fixed number of rounds would not guarantee the correct computation. 
%Consequently, rather than running each phase for a fixed number of communication rounds, we do it until a condition suited for each phase is violated (cf. Algorithm~\ref{simprotocol}), and we count the number of communication rounds to complete the computation.

%As expected, the synchronization and bookkeeping has to be centralized rather than distributed, and nodes are assigned labels for reference. Nevertheless, the running times obtained may be used to fix the number of rounds needed to run the protocol distributedly in a practical \ADN, should the input behavior be known. %Moreover, should the topology dynamics be stochastic and known, the described approach yields a Montecarlo protocol. 

In the following paragraphs, we provide further details on the changes applied to each phase of \IC, and how each phase is implemented in our simulator. %the implementation of each phase in our simulator.

During the collection phase of \IC non-leader nodes are initially assigned a unit of energy, which is later disseminated towards the leader using a gossip-based approach~\cite{KDGgossip,flowupdate,FMT:aggJournal}. That is, each non-leader node repeatedly shares a fraction of its energy with each neighbor. Given that the leader keeps all the energy received, it eventually collects most of the energy in the system. In the original \IC protocol, for each candidate size $k$, the number of rounds for sharing energy is fixed to a function $\tau(k)$ that has not be proven to be sub-exponential in the worst case. Thus, to evaluate whether in practice a polynomial number of rounds is enough, in our simulations we iterate the energy transfer until the conditions needed for the verification phase are met. (Refer to Lines~\ref{collbegin} to~\ref{collend} in Algorithm~\ref{simprotocol}.) That is, until the leader 
%has too much energy for the current candidate size (if the guess $k$ of the leader is the correct size of the system, the total energy is $k-1$) or the leader 
has collected an amount of energy such that, if its guess is correct, non-leader nodes have transferred almost all their energy, i.e. all non-leader nodes have residual energy smaller than or equal to $1/k^{1.01}$.

In the \ADN model communication is carried out in rounds. In each round, every node receives from each neighboring node simultaneously. To simulate this exchange in the collection phase we follow the approach used to analyze gossip-based protocols~\cite{KDGgossip,flowupdate,FMT:aggJournal}. That is, the energy sharing process is simulated by a multiplication of the vector of energies by a matrix of fractions shared. 
More precisely, we maintain a vector $\mathbf{e_r}=\big(e_{r_1},e_{r_2},\dots,e_{r_n}\big)$ of energies, where $e_{r_i}$ is the energy of node $i$ at the beginning of round $r$. Additionally, for each input graph with set of links $E(r)$, we define a matrix $\mathbf{F}(E(r))=\big(f_{ij}\big)$, where $f_{ij}$ is the fraction of energy of node $j$ that node $i$ receives. 
We will refer to this matrix as $\mathbf{F}(E)$ when the round number is clear from context.
For example, for a complete graph with set of links $E$ where node $1$ is the leader it is
\begin{align}
\mathbf{F}(E) &=
\left( \begin{array}{ccccc}
1 & \frac{1}{2\Delta} & \frac{1}{2\Delta} & \frac{1}{2\Delta} & \ldots \\
0 & 1-\frac{n-1}{2\Delta} & \frac{1}{2\Delta} & \frac{1}{2\Delta} & \ldots \\
0 & \frac{1}{2\Delta} & 1-\frac{n-1}{2\Delta} & \frac{1}{2\Delta} & \ldots \\
0 & \frac{1}{2\Delta} & \frac{1}{2\Delta} & 1-\frac{n-1}{2\Delta} & \ldots \\
\vdots & \vdots & \vdots &\vdots & \ddots 
\end{array} \right)\label{eqF}
\end{align}
Then, the energy of nodes is iteratively computed as $\mathbf{e_{r+1}}=\mathbf{F \cdot e_r}^T$. Notice that the matrix $\mathbf{F}$ has to be changed each time the input graph changes.

%That is, the energy sharing process in the collection phase is simulated as a multiplication of a matrix of fractions shared with neighbors by the vector of energies. \mig{This procedure guarantees that the energies are shared concurrently among nodes as in \IC. Otherwise, if energy was updated for one node at a time, the whole process could potentially be faster.}

During the verification phase of \IC non-leader nodes disseminate towards the leader the value of the maximum energy held by any non-leader node. If the residual energy of some node is greater than the above threshold, the current candidate size is deemed incorrect by the leader. To guarantee that the leader receives from all nodes, all non-leader nodes iteratively broadcast and update the maximum energy heard, starting from their own. This phase does not tolerate disconnection of the network, since then some nodes might not be heard by the end of the loop. To evaluate disconnected topologies, in our simulations we continue the iteration until  the leader has received from all nodes. (Refer to Lines~\ref{verifbegin} to~\ref{verifend} in Algorithm~\ref{simprotocol}.)

If the candidate size was found to be correct in the verification phase, a halting message is broadcast throughout the network in the notification phase. To synchronize the computation, the notification phase of \IC runs for a fixed number of rounds, independently of whether the current candidate size is correct or not. As with the verification phase, the notification phase does not tolerate disconnection of the network. To evaluate disconnected topologies, in our simulations we continue the iteration also for the notification phase, in this case until all nodes receive from the leader. (Refer to Lines~\ref{notifbegin} to~\ref{notifend} in Algorithm~\ref{simprotocol}.)

To handle disconnection, the approach followed in the latter two phases is centralized. Nevertheless, the running times obtained are not better than in \IC, since the verification and notification phases run for at least the same number of rounds.

In our simulations the synchronization and bookkeeping are centralized rather than distributed, and nodes are assigned labels for reference. Nevertheless, the running times obtained may be used to fix the number of rounds needed to run the protocol distributedly in a practical \ADN, should the input behavior be known.

%\onecolumn
\begin{algorithm}[htbp]
\DontPrintSemicolon
\SetKwData{true}{true}
\SetKwData{false}{false}
\SetKwFor{collection}{Collection Phase:}{}{end phase}
\SetKwFor{verification}{Verification Phase:}{}{end phase}
\SetKwFor{notification}{Notification Phase:}{}{end phase}
$k\leftarrow 1$,
$IsCorrect\leftarrow \false$,
$r\leftarrow 1$,
$E\leftarrow$ new set of links\;
\While{$\neg IsCorrect$}{
	$k\leftarrow k+1$,
	$IsCorrect \leftarrow \true$\;

	\BlankLine
	\BlankLine
	%\tcp{Collection Phase}
	\collection{\label{collbegin}}{
		$\big(e_1,e_2,e_3,\dots,e_n\big)\leftarrow\big(0,1,1,\dots,1\big)$\tcp*{vector of energies}
%		\While{$e_1\leq k-1$ {\bf and} $e_1<\ale{k-1-1/k^{1.01}}$}{%n-1-1/n^{1.01}$}{
		\While{$e_1 < k-1-1/k^{1.01}$}{
			$\big(e_1,e_2,\dots,e_n\big) \leftarrow \mathbf{F}(E) \cdot \big(e_1,e_2,\dots,e_n\big)^T$\tcp*{broadcast simulation}
			\lIf{$r\equiv 0\mod T$}{$E\leftarrow$ new set of links}
			$r \leftarrow r +1$\;
			\label{collend}
		}
	}
	
	\BlankLine
	\BlankLine
	%\tcp{Verification Phase}
	\verification{\label{verifbegin}}{
	\lIf{$e_1> k-1$}{$IsCorrect \leftarrow \false$}
	$\big(e'_1,e'_2,e'_3,\dots,e'_n\big)\leftarrow\big(0,e_2,e_3\dots,e_n\big)$\tcp*{vector of max energy heard}
	\For{$1+\lceil k/(1-1/k^{1.01}) \rceil$ iterations}{ 
		\lFor(\tcp*[f]{broadcast simulation}){each $i$}{$e''_i \leftarrow \max_{j\in N(i,E)\cup\{i\}} e'_j$}
		$\big(e'_1,e'_2,\dots,e'_n\big)\leftarrow\big(e''_1,e''_2,\dots,e''_n\big)$\;
			\lIf{$r\equiv 0\mod T$}{$E\leftarrow$ new set of links}
			$r \leftarrow r +1$\;
	}
\textcolor{light-gray}{
\tcp{for $G_{n,p}$ input only:}
	\While{not all nodes ``heard'' by leader}{ 
		\lFor(\tcp*[f]{broadcast simulation}){each $i$}{$e''_i \leftarrow \max_{j\in N(i,E)\cup\{i\}} e'_j$}
		$\big(e'_1,e'_2,\dots,e'_n\big)\leftarrow\big(e''_1,e''_2,\dots,e''_n\big)$\;
			\lIf{$r\equiv 0\mod T$}{$E\leftarrow$ new set of links}
			$r \leftarrow r +1$\;
	}
}
	\lIf{$e'_1>1/k^{1.01}$}{$IsCorrect \leftarrow \false$\label{verifend}}
	}
	
	\BlankLine
	\BlankLine
	%\tcp{Notification Phase}
	\notification{\label{notifbegin}}{
	$\big(h_1,h_2,h_3,\dots,h_n\big)\leftarrow\big(IsCorrect,\false,\false,\dots,
	\false\big)$\tcp*{vector of halt flags}
	\For{$k$ iterations} { 
		\lFor(\tcp*[f]{broadcast simulation}){each $i$}{$h'_i \leftarrow \bigvee_{j\in N(i,E)} h_j$}
		$\big(h_1,h_2,\dots,h_n\big)\leftarrow\big(h'_1,h'_2,\dots,h'_n\big)$\;
			\lIf{$r\equiv 0\mod T$}{$E\leftarrow$ new set of links}
			$r \leftarrow r +1$\;
	}
\textcolor{light-gray}{
\tcp{for $G_{n,p}$ input only:}
	\While{$\lnot (h_1 \Rightarrow \bigwedge_{i\in V} h_i)$} { 
		\lFor(\tcp*[f]{broadcast simulation}){each $i$}{$h'_i \leftarrow \bigvee_{j\in N(i,E)} h_j$}
		$\big(h_1,h_2,\dots,h_n\big)\leftarrow\big(h'_1,h'_2,\dots,h'_n\big)$\;
			\lIf{$r\equiv 0\mod T$}{$E\leftarrow$ new set of links}
			$r \leftarrow r +1$\;
	}
}
	\label{notifend}
	}
}

	\BlankLine
	\BlankLine
output $k$\;
\caption{Centralized simulation of leader and non-leader nodes running \IC~\cite{opodisCounting} on $T$-stable topologies. 
%(Included  in the Appendix as Algorithms~\ref{algo:leader} and~\ref{algo:no-leader} respectively for self-containment.) 
$V$ is the set of nodes and $E$ is the set of links.
%A new set of links $E$ is chosen every $T$ rounds of communication (cf. Section~\ref{sec:input}).
Node $1$ is the leader node. 
The other node labels are used for reference only, but not used to make decisions. 
%According to the $T$-stable model tested, a new topology is tested every $T$ rounds.
%Topology changes are not shown in the algorithm. 
$N(i,E)$ is the set of neighbors of node $i$ according to $E$. 
$\mathbf{F}(E)$ is the matrix of fractions shared according to $E$ (cf. Equation~\ref{eqF}).} 
\label{simprotocol}
\end{algorithm}

\section{Input Topologies}
\label{sec:input}

To thoroughly evaluate the performance of \IC, we have produced different topologies that may appear in practice. The parameters of our simulations include the size of the network $n$, the upper bound on the number of neighbors $\Delta$, and the period of stable topology $T$. 

We evaluate random tree topologies rooted at the leader node, drawn uniformly as described below. As extremal cases of a tree topology, we also evaluated a star rooted at the leader node and a path where the leader is the end point.
We also consider Erdos-Renyi random graphs, for which we additionally parameterize the probability $p$ that any given pair of nodes are neighbors. 
Although graphs have better conductance than the trees underlying them, and consequently graphs achieve convergence faster for gossip-based protocols~\cite{SJ:count}, we evaluate the latter inputs for consistency with previous works. 
Random graphs may be disconnected, but in our simulator \IC has been modified to handle disconnected components. (Refer to Section~\ref{sec:protocols} for details.)

We evaluated the performance of \IC for all values of $n\in [3,75]$,
and $T\in\{1,10,20,40,80,160,320,640,1280,\infty\}$, where $T=\infty$ corresponds to a static network. 
\mig{For the star and path topologies, we permute the labels every $T$ rounds, since no other change is possible given that the topology is fixed.
On the other hand, for random trees and random graphs, we produce a whole new network every $T$ rounds, since the topology is less restricted (trees) or it is not restricted at all (graphs).
}
For stars and random graphs, we fixed the degree upper bound $\Delta=n-1$. 
For the former because that is the maximum degree and for the latter to guarantee a uniform draw.
For random trees and paths we used $\Delta=2^i$, for all values of $i>0$ such that $2^i\leq n-1$. 
For random graphs we evaluated $p\in\{0.1,0.2,0.3,0.4,0.5\}$.
All our results where computed as the average over $100$ executions of the protocol.

To produce our random rooted unlabeled trees we used the algorithm RANRUT presented in~\cite{combalgs}, which was proved to provide a uniform distribution on the equivalence classes defined by isomorphisms.
The tree so obtained may have maximum degree larger than $\Delta$. When that is the case, we prune the tree moving subtrees downwards until all nodes have at most $\Delta$ neighbors. This procedure increases (or does not change) the longest path to the leader, which increases (or does not change) the running time of \IC. Thus, with respect to a uniform distribution on rooted unlabeled trees of maximum degree $\Delta$, our input distribution is biased ``against'' \IC providing stronger guarantees.

In contrast, topologies obtained connecting pairs of nodes stochastically until the graph is connected, usually have relatively low diameter, which may speed up the energy transfer. 
Algorithm~\ref{alg:ranrut} summarizes the rooted tree network generator used in our simulations.

%\onecolumn
\begin{algorithm}
\DontPrintSemicolon
\SetKwData{tree}{tree}
\SetKwFunction{GENTREE}{GENTREE}
\SetKwFunction{SIZES}{SIZES}
\SetKwFunction{DISTRIB}{DISTRIB}
\SetKwFunction{RANRUT}{RANRUT}
\SetKwFunction{PRUNE}{PRUNE}
\SetKwFunction{RECPRUNE}{RECPRUNE}
\SetKwFunction{ATTACH}{ATTACH}
\SetKwFunction{subtrees}{subtrees}
\SetKwFunction{root}{root}
\SetKwProg{function}{Function}{}{}
\SetKwProg{proc}{Procedure}{}{}
\function{\GENTREE{$n,\Delta$}}{
$t\leftarrow \SIZES(n)$
\tcp{Compute number of unlabeled rooted trees of size $1,2,\dots,n$.}
$p\leftarrow \DISTRIB(t,n)$
\tcp{Compute distributions on subtrees for each $n$.}
$\tree\leftarrow \RANRUT(p,n)$
\tcp{Choose an unlabeled rooted tree uniformly at random.}
\PRUNE($\tree,\Delta$)
\tcp{Move subtrees downwards until max degree of \tree is $\Delta$.}
\Return{\tree}
}

\caption{Random tree generator algorithm. Auxiliary functions in Algorithm~\ref{alg:aux}. %in the Appendix.
\label{alg:ranrut}
%root(t): root node of tree t.
%subtrees(i): set of subtrees of tree node i.
}
\end{algorithm}
%\twocolumn

\begin{algorithm}
\DontPrintSemicolon
\SetKwData{tree}{tree}
\SetKwFunction{GENTREE}{GENTREE}
\SetKwFunction{SIZES}{SIZES}
\SetKwFunction{DISTRIB}{DISTRIB}
\SetKwFunction{RANRUT}{RANRUT}
\SetKwFunction{PRUNE}{PRUNE}
\SetKwFunction{RECPRUNE}{RECPRUNE}
\SetKwFunction{ATTACH}{ATTACH}
\SetKwFunction{subtrees}{subtrees}
\SetKwFunction{root}{root}
\SetKwProg{function}{Function}{}{}
\SetKwProg{proc}{Procedure}{}{}
\SetKwData{t}{t}
\SetKwData{p}{p}
\BlankLine
\function{\SIZES{$n$}}{
Let $\t[1\dots n]$ be a new array\;
$\t[1]\leftarrow 1$, $\t[2]\leftarrow 1$\;
%\For(\tcp*[f]{for each value of $i>2$ set $t_i\leftarrow\sum_{j=1}^\infty\sum_{d=1}^\infty \frac{dt_{i-jd}t_d}{i-1}$}){$i=3$ to $n$}
\For{$i=3$ to $n$}
{
\tcp{$t_i\leftarrow\sum_{j=1}^\infty\sum_{d=1}^\infty \frac{dt_{i-jd}t_d}{i-1}$}
$\t[i]\leftarrow 0$\;
\For{$j=1$ to $n$}{
\For{$d=1$ to $n$}{
\lIf{$jd<i$}{$\t[i]\leftarrow \t[i]+d\cdot \t[i-jd]\cdot \t[d]$}
}
}
$\t[i]\leftarrow \t[i]/(i-1)$\;
}
\Return{\t}
}
\BlankLine
\function{\DISTRIB{$\t,n$}}{
%\For(\tcp*[f]{for each value of $k>2$ set $p_{j,d}\leftarrow\frac{dt_{k-jd}t_d}{(k-1)t_k}$}){$k=3$ to $n$}
Let $\p[1\dots n][1\dots n][1\dots n]$ be a new 3D array\;
\For{$k=3$ to $n$}
{
\For{$j=1$ to $n$}
{
\For{$d=1$ to $n$}
{
\lIf{$jd<k$}{$\p[k][j][d]\leftarrow d \cdot \t[k-jd]\cdot\t[d] / ((k-1)\cdot\t[k])$}
\lElse{$\p[k][j][d]\leftarrow0$}
}
}
}
\Return{\p}
}
\BlankLine
\function{\RANRUT{$\p,n$}}{
\lIf{$n\leq2$}{$\tree\leftarrow$new tree of size $n$}
%\Else(\tcp*[f]{$n>2$}){
\Else{
with probability distribution $\p[n][j][d]$ draw a pair $(j,d)$\;
$\tree\leftarrow\RANRUT(\p,n-jd)$\;
\For{$j$ times}{
$\tree'\leftarrow\RANRUT(\p,d)$\;
attach $\tree'$ to the root of \tree\;
}
}
\Return{$\tree$\;}
}
\BlankLine
\proc{\PRUNE{$\tree,\Delta$}}{
\While{root of \tree has more than $\Delta$ neighbors}{ 
%move the rightmost subtree of the root of \tree to a leaf\;
detach the rightmost subtree \tree' from the root of \tree\;
\ATTACH($\tree,\tree',\Delta$)\;
}
%\ForEach{$\tree'\in\subtrees(\root(\tree))$}{
\ForEach{subtree \tree' of the root of \tree}{
%\RECPRUNE($\tree',\Delta$)\;
\PRUNE($\tree',\Delta$)\;
}
}
%
%\BlankLine
%%\setcounter{AlgoLine}{0}
%\proc{\RECPRUNE{$\tree,\Delta$}}{
%\While(\tcp*[f]{one less due to link to parent}){root of \tree has more than $\Delta-1$ children}{ 
%move the rightmost subtree of the root of \tree to a leaf\;
%}
%%\ForEach{$\tree'\in\subtrees(\root(\tree))$}{
%\ForEach{subtree \tree' of the root of \tree}{
%\RECPRUNE($\tree',\Delta$)\;
%}
%}
%
\BlankLine
\proc{\ATTACH{$\tree,\tree',\Delta$}}{
\If{root of \tree has less than $\Delta$ neighbors}{
attach $\tree'$ to the root of \tree\;
}
\Else{
choose a subtree \tree'' of the root of \tree uniformly at random\;
\ATTACH($\tree'',\tree',\Delta$)\;
}
}
\caption{Auxiliary functions for Algorithm~\ref{alg:ranrut}. Refer to~\cite{combalgs} for details on \texttt{SIZES}, \texttt{DISTRIB}, and \texttt{RANRUT}.} 
\label{alg:aux}
%root(t): root node of tree t.
%subtrees(i): set of subtrees of tree node i.
\end{algorithm}
%\twocolumn

\section{Simulation Platform}
\label{sec:platform}

We developed our own simulator and input generator implementing Algorithms~\ref{simprotocol} and~\ref{alg:ranrut} in Java 8. 
The simulations were carried out on a cluster facility at Kean University known as Puma. Puma is a 130 node, 1040 core Dell cluster running Red Hat Enterprise Linux and Rocks+ 4.3. Each node has 2 quad core 2.6 GHz Xeon processors, 16GB RAM, and 350GB local storage, and are connected with Gigabit ethernet. 

Notice that our simulator is centralized. That is, rather than using the cluster to simulate the communication among nodes, we simply run various instances of the simulator in parallel to speedup our experiments. Using the cluster to simulate the communication among nodes would not have provided any additional insight, given that in the \ADN model communication is reliable and our protocol is deterministic.

%!TEX root = main_llncs.tex

\section{Discussion}
\label{sec:discussion}

%The problem of Counting in \ADNs is challenging because lack of identifiers and changing topology make difficult to decide if a new message has been received before from the same node. Counting protocols~\cite{opodisCounting,conscious,oracle} overcome this challenge checking candidate sizes iteratively until the correct size is obtained. However, the worst-case convergence time of the gossip-based process used in those protocols has been bounded only exponentially, yielding only exponential guarantees for \IC~\cite{opodisCounting}. The situation is even worse in other works where a huge overestimate (proved in~\cite{spirakis}) is initially used as the candidate size, yielding a double-exponential protocol~\cite{conscious}.
%Other Counting protocols do not terminate~\cite{conscious}, 
%or terminate but do not provide running-time guarantees~\cite{oracle},
%or compute only an exponential upper bound on the network size~\cite{spirakis}. 
%Other heuristic protocols do not guarantee that the computation is correct~\cite{experimentalConscious}.
%
%The main goal of this work was to show that in practice \IC may compute the size of \ADNs in a polynomial number of rounds of communication, obtaining always the correct result. Nevertheless, the extensive simulations carried out provided also very interesting observations about the impact of topology and dynamics in performance. The details follow.

\begin{figure*}[t]
       \centering
       \large
%       \subfloat[Number of correct rounds.][]
       {\label{fig:tree}\resizebox{.5\textwidth}{!}{\input{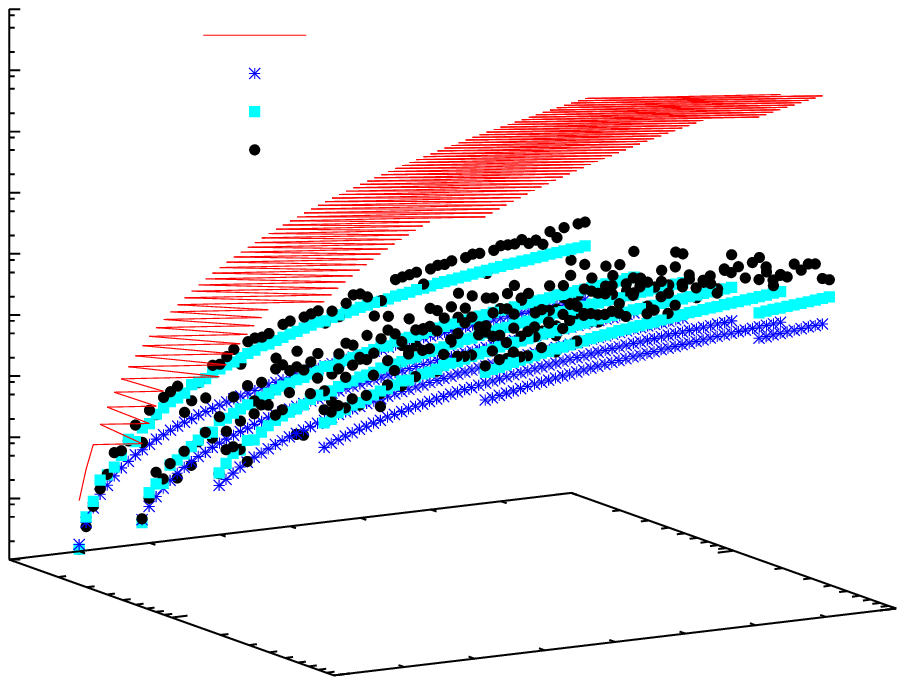}}}%
%       \qquad
%       \subfloat[Cumulative master utility.][]
       {\label{fig:path}\resizebox{.5\textwidth}{!}{\input{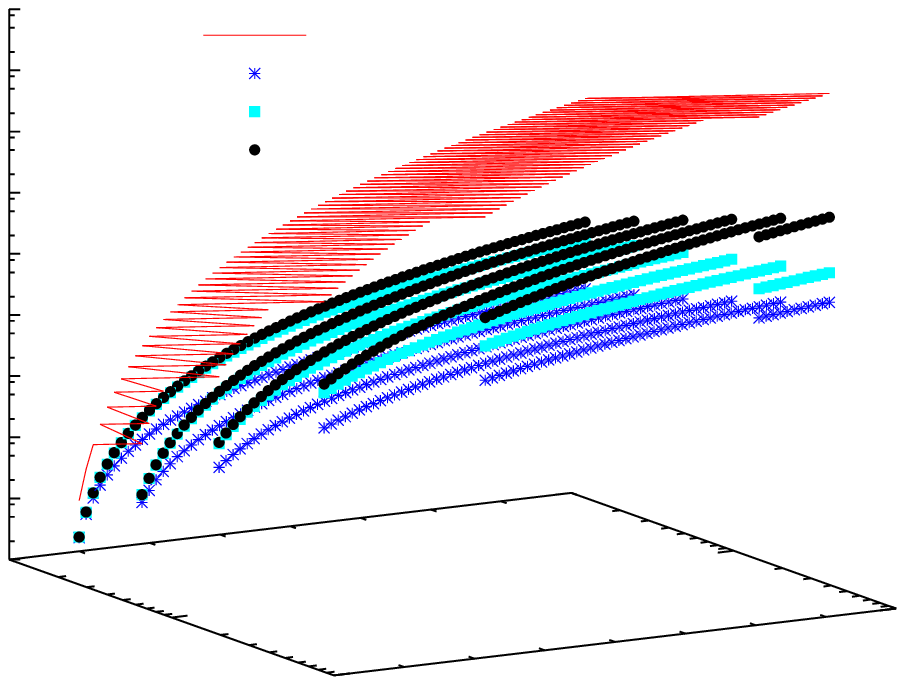}}}\\
       \vspace{.2in}
%       \subfloat[Cumulative follower worker utility.][]
       {\label{fig:star}\resizebox{.5\textwidth}{!}{\input{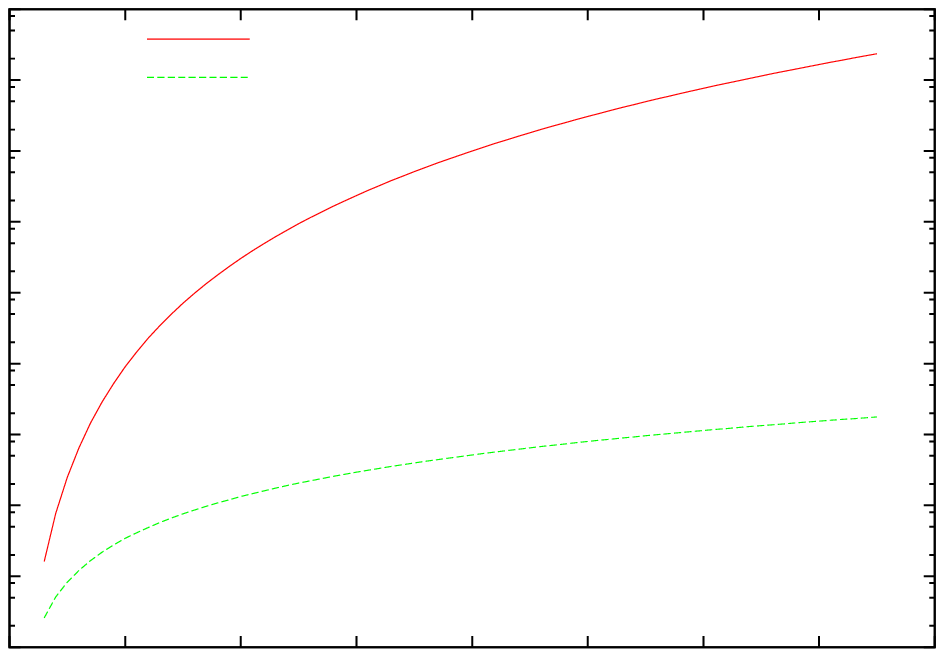}}}%
%       \qquad
%       \subfloat[Rounds to detection/convergence.][]
       {\label{fig:gnp}\resizebox{.5\textwidth}{!}{\input{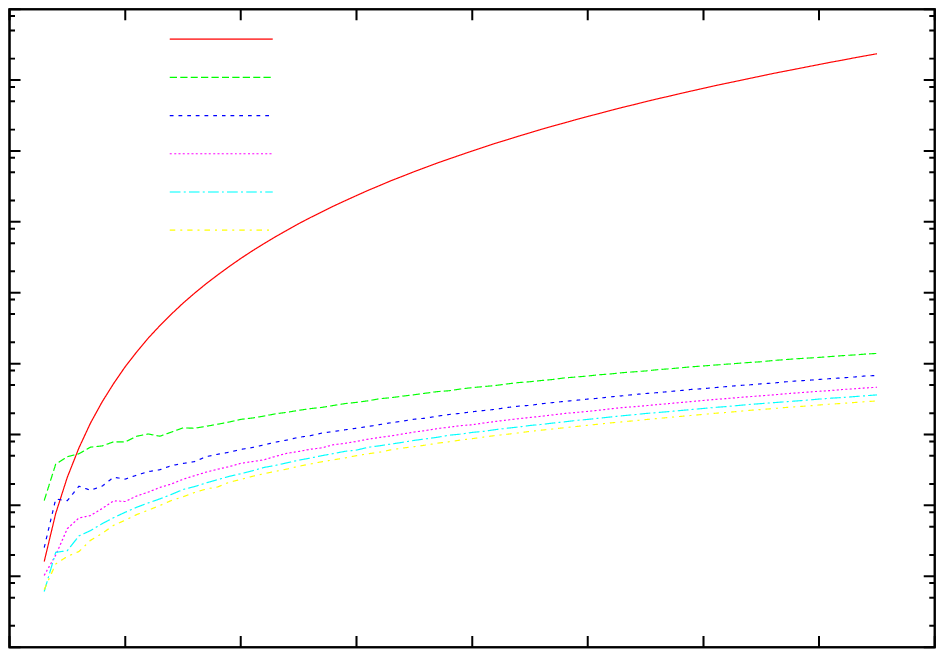}}}%
       \caption[]{\IC time performance compared with a polynomial function.}%
       \label{fig:polynomial}%
\end{figure*}

For all the topologies and parameter combinations evaluated, \IC has proved to be polynomial. We plot here only a subset of our results for succinctness. Consider for instance Figure~\ref{fig:polynomial}, where we plot the number of rounds to complete the computation (log scale) as a function of the network size, for various values of degree upper bound $\Delta$, interval of stability $T$, and probability $p$ of being connected in the random graph. We also plot the function $\Delta n^4$ to contrast the growth of such polynomial function with the results obtained. It can be seen that all our results indicate a rate of growth asymptotically smaller than $\Delta n^4$. (The upper bound could be tightened but we choose a loose one for clarity.)
For small network sizes, random graphs introduce additional delays due to disconnection, but as the network scales the dynamic topology overcomes the effect of disconnections.  

\begin{figure*}[t]
       \centering
       \large
%       \subfloat[Number of correct rounds.][]
       {\label{fig:treeDeltaT10}\resizebox{.5\textwidth}{!}{\input{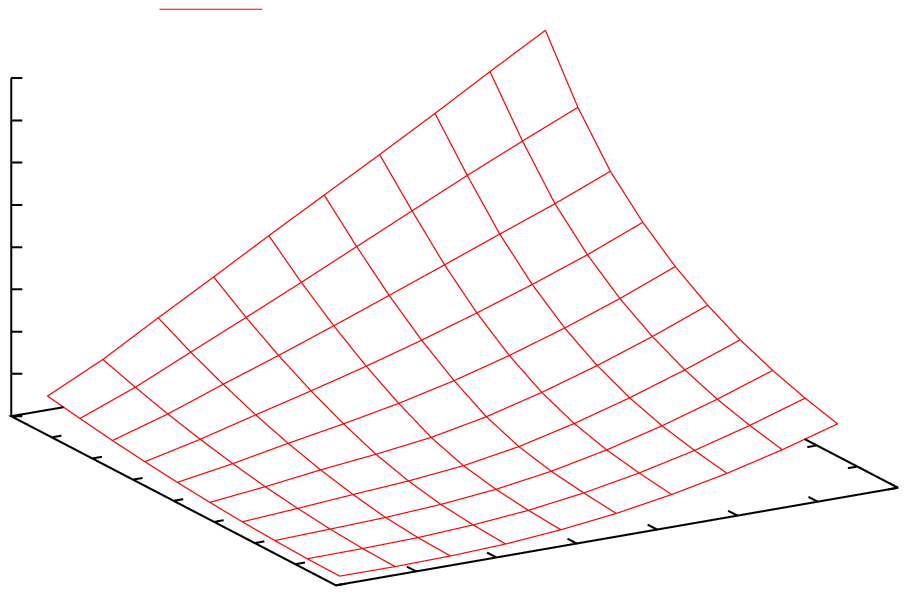}}}%
%       \qquad
%       \subfloat[Cumulative master utility.][]
       {\label{fig:pathDeltaT10}\resizebox{.5\textwidth}{!}{\input{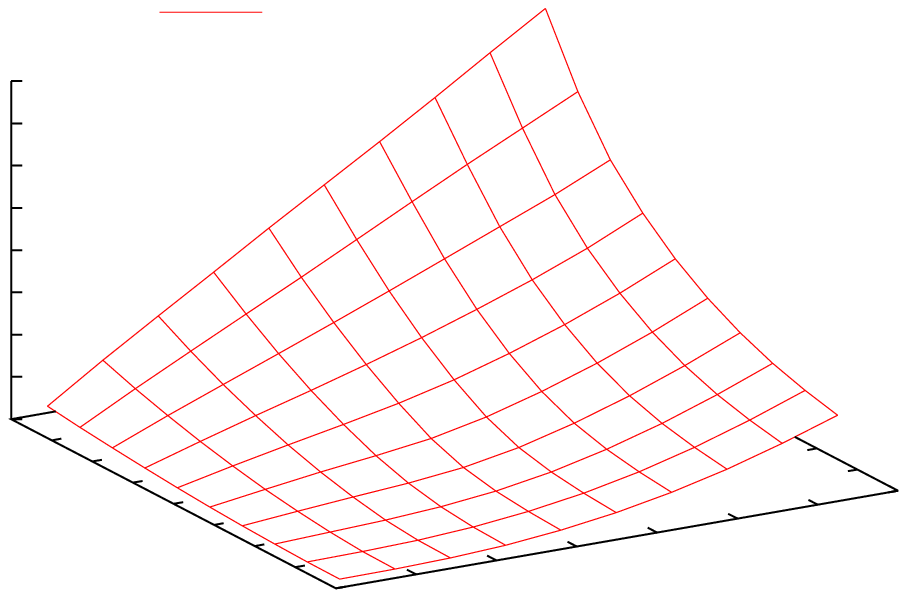}}}\\
       \vspace{-.2in}
%       \subfloat[Number of correct rounds.][]
       {\label{fig:treeDeltaT1280}\resizebox{.5\textwidth}{!}{\input{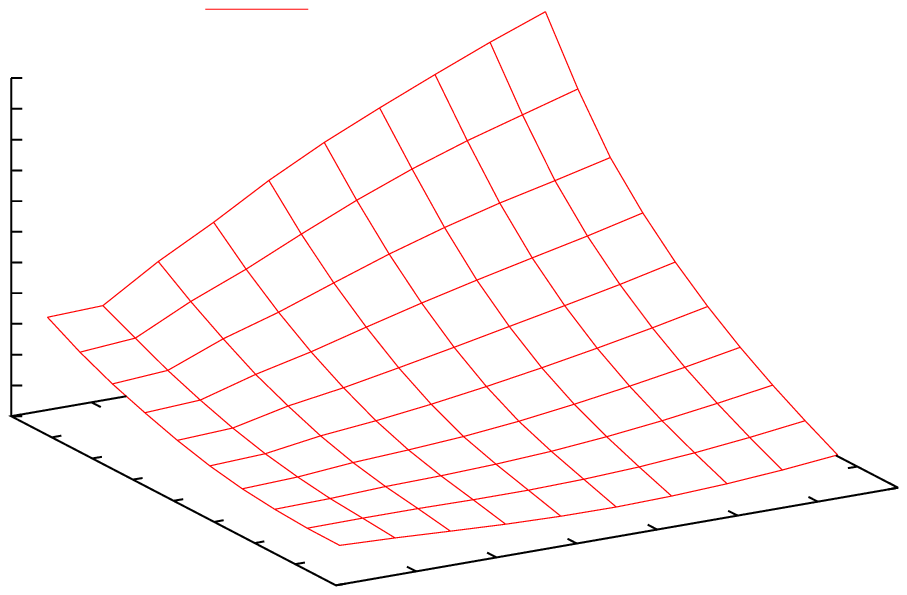}}}%
%       \qquad
%       \subfloat[Cumulative master utility.][]
       {\label{fig:pathDeltaT1280}\resizebox{.5\textwidth}{!}{\input{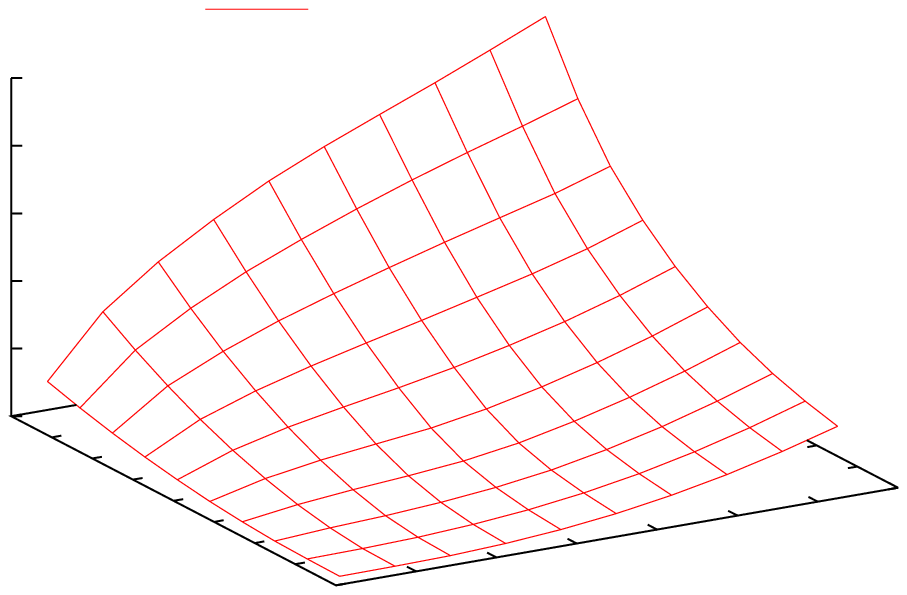}}}
       \caption[]{\IC time performance vs. degree upper bound.}%
       \label{fig:delta}%
\end{figure*}

For the tree and path topologies the degree upper bound $\Delta$ was checked for various values. The performance of the protocol with respect to $\Delta$ for those inputs is illustrated in Figure~\ref{fig:delta}. For paths, the running time increases with $\Delta$. This is expected given that, although nodes change positions, the topology is always a path. $\Delta$ is an upper bound, hence it may be increased, but the only impact is a reduction on the fraction of energy shared with neighbors (see Algorithm~\ref{simprotocol}). Such reduction applies to the energy transfered towards the leader as well as away from the leader. Nevertheless, the balance yields a slower dissemination towards the leader in the collection phase.
The same behavior can be observed for the tree topology. This is somewhat surprising because we would expect a larger $\Delta$ to yield a shallower random tree. However, the impact of a smaller fraction of energy shared dominates. Notice also in Figure~\ref{fig:delta} that these observations apply whether the topology changes frequently or not.

\begin{figure*}[t]
       \centering
       \large
%       \subfloat[Number of correct rounds.][]
       {\label{fig:10IntGnp}\resizebox{.5\textwidth}{!}{\input{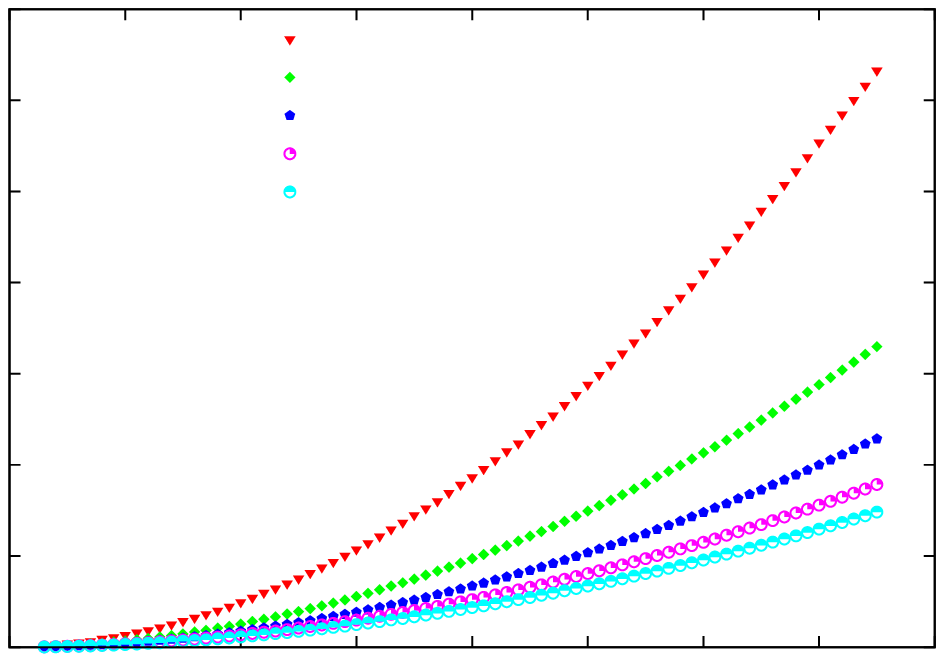}}}%
%       \qquad
%       \subfloat[Cumulative master utility.][]
%       {\label{fig:80IntGnp}\resizebox{.5\textwidth}{!}{\input{./plots75nodes/plot_80IntGnp}}}\\
%       \subfloat[Cumulative follower worker utility.][]
%       {\label{fig:160IntGnp}\resizebox{.5\textwidth}{!}{\input{./plots75nodes/plot_160IntGnp}}}%
%       \qquad
%       \subfloat[Rounds to detection/convergence.][]
       {\label{fig:320IntGnp}\resizebox{.5\textwidth}{!}{\input{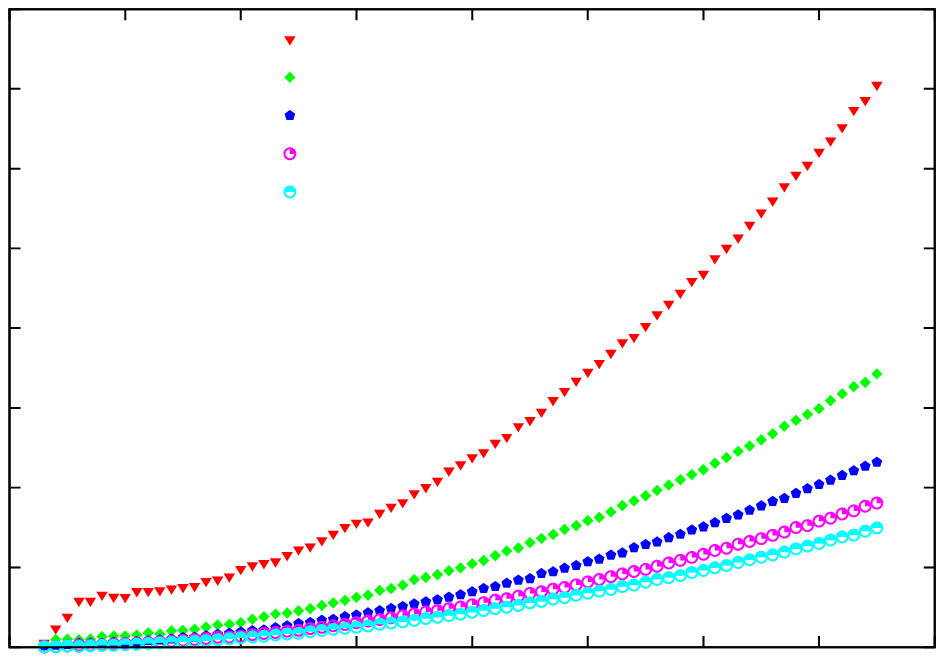}}}
       \caption[]{\IC time performance on random graphs for different intervals of stability.}%
       \label{fig:intervalGnp}%
\end{figure*}

Figure~\ref{fig:intervalGnp} shows the performance of the protocol for random graph inputs. As expected, when the probability $p$ of any given pair of nodes being connected is low, the time needed to complete the computation grows drastically if the network does not change frequently, because some nodes cannot reach the leader. Indeed, even for $10$-stable networks the performance when $p=0.1$ is much worse than for higher values of $p$ when the network is likely to be connected. But if we consider less-dynamic networks, such as $320$-stable, the performance for $p=0.1$ is much worse than others even for small networks.

\begin{figure*}[t]
       \centering
       \large
%       \subfloat[Number of correct rounds.][]
       {\label{fig:10Int}\resizebox{.5\textwidth}{!}{\input{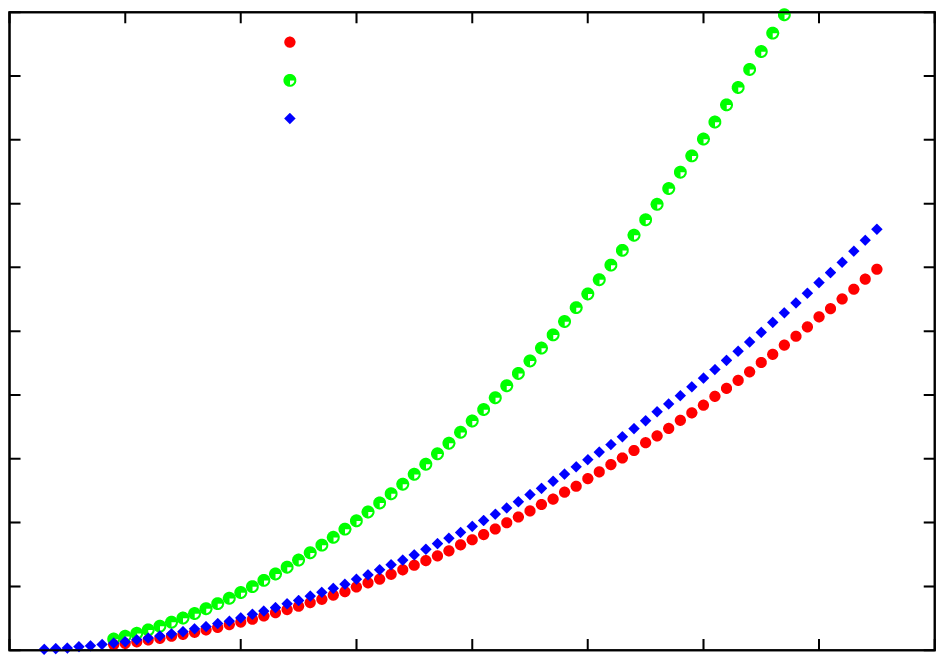}}}%
%       \qquad
%       \subfloat[Cumulative master utility.][]
       {\label{fig:160Int}\resizebox{.5\textwidth}{!}{\input{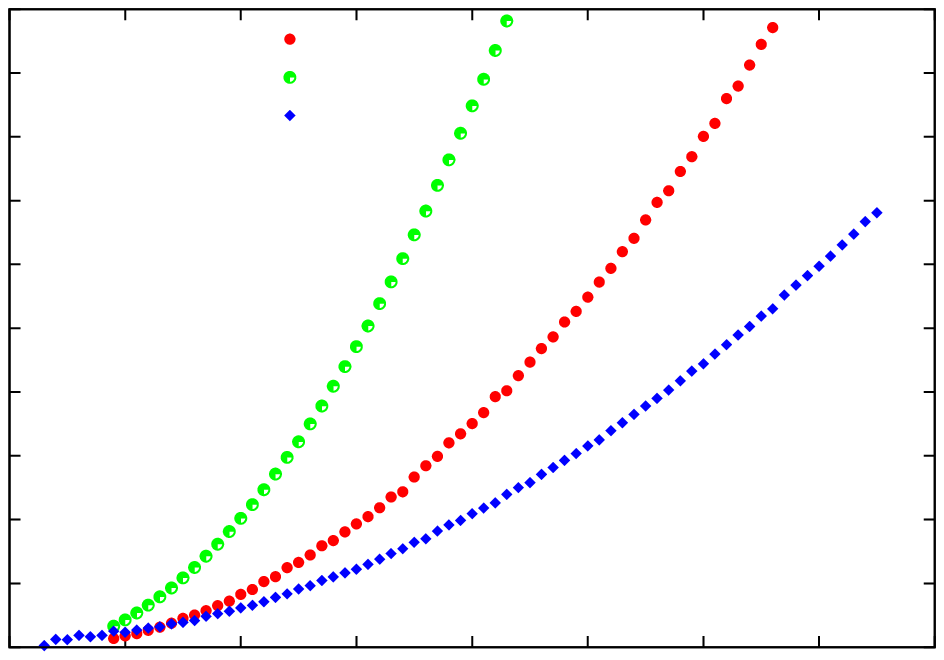}}}\\
%       \subfloat[Cumulative follower worker utility.][]
       {\label{fig:1280Int}\resizebox{.5\textwidth}{!}{\input{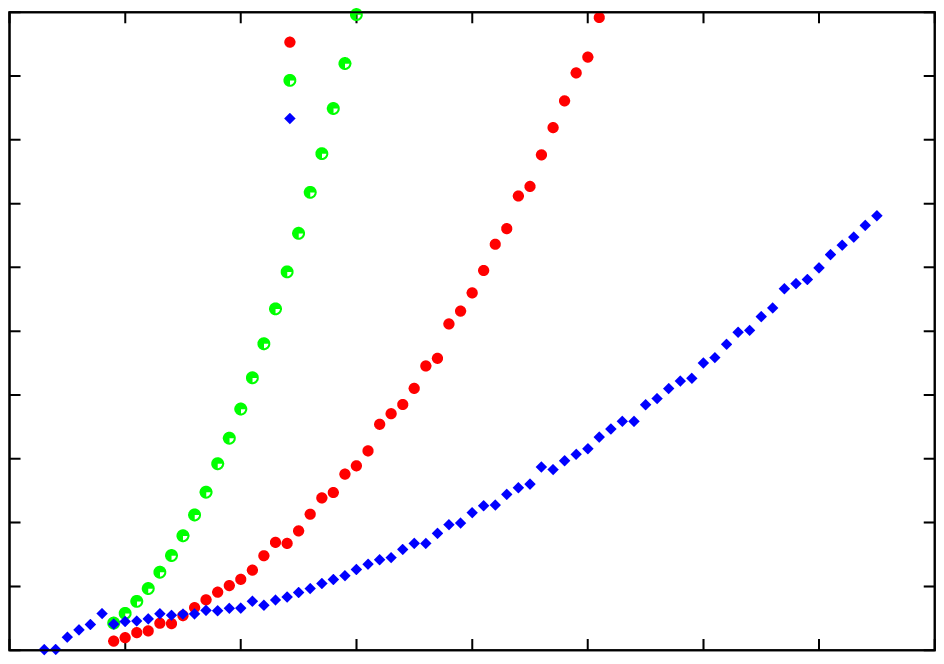}}}%
%       \qquad
%       \subfloat[Rounds to detection/convergence.][]
       {\label{fig:static}\resizebox{.5\textwidth}{!}{\input{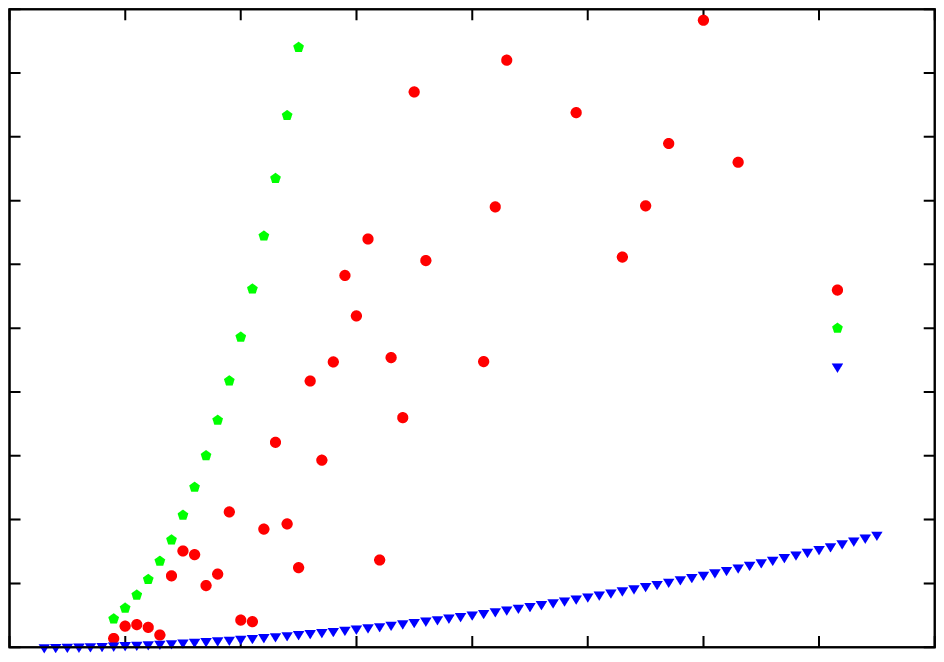}}}
       \caption[]{\IC time performance for different intervals of stability.}%
       \label{fig:interval}%
\end{figure*}

Figure~\ref{fig:interval} illustrates the performance for different intervals of stability and representative cases of each type of input. It can be observed that when the network is highly dynamic ($10$-stable) the random tree and random graph produce similar results. To understand why, consider for instance $n=25$. Changing the topology every 10 rounds yields a network where, on average over the roughly 8000 rounds, every pair of nodes is connected. The difference between tree and graph becomes bigger as the topology changes less frequently, when the impact of the lower conductance of the tree becomes dominant. 

For static random trees, we can observe also in Figure~\ref{fig:interval} that the running times varied significantly from one network size to another. This is due to the particular trees that happened to be drawn, which sometimes have better conductance than others. 
The random graph was not evaluated for static networks since permanent disconnection would stop the protocol from completing the computation. On the other hand, for the star input the only dynamics introduced was a permutation of labels, which are not used by \IC and therefore has no impact on the running time. Thus, we illustrate the results only in the static topology plots. As expected, a star yields the fastest running time since all nodes are connected to the leader.

Finally, Figure~\ref{fig:interval} shows also the performance on a path input. Given that the topology is fixed to a path, as with the star topology, the only dynamics introduced is a permutation of labels. Then, as expected, the running time increases significantly as the network changes less frequently. The reason being that in a path where most of the nodes do not change much their distance to the leader, the fraction of energy received by the leader in each round from distant nodes is inverse exponential on that distance.

\begin{figure}[htbp]
       \centering
       \large
       \resizebox{.5\textwidth}{!}{\input{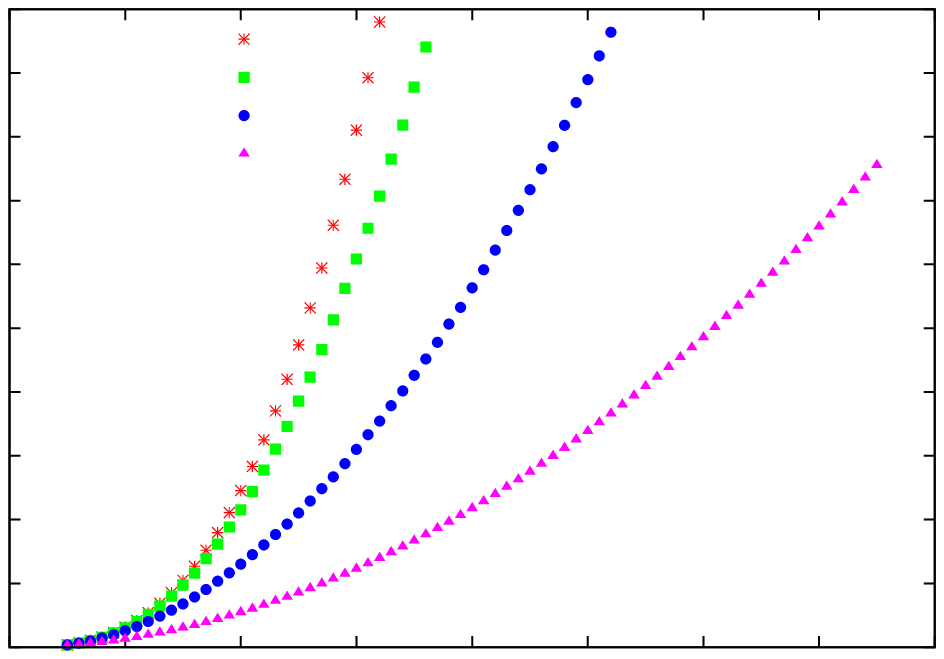}}
       \caption[]{\IC time performance for a path topology.}%
       \label{fig:pathByInterval}%
\end{figure}

In summary, our simulations have shown that \IC runs in polynomial time for all the networks evaluated. 
The set of inputs tested comprise average topologies likely to appear in practice as well as extremal cases.
%Given that the set of inputs tested comprise average topologies likely to appear in practice as well as extremal cases, we conjecture that \IC runs in polynomial time even in the worst case. A formal proof of such conjecture remains open, but the intuition can be obtained as follows.
%Consider the case of a network with a static path topology. 
Our simulations showed the static path to have the worse time performance among all inputs tested.
(For illustration, see  in Figure~\ref{fig:pathByInterval} the case of a path with $\Delta=4$ for various values of $T$.)
\miguel{Intuition on why \IC would be polynomial in static paths can be obtained as follows.} 
To bound the collection time, consider the inverse gossip-based process of computing the average starting with the leader having energy $n$ and all other nodes $0$. This process can be modeled as a Markov chain, whose convergence time bounds the time to compute the average (hence, the collection time). It is known~\cite{levin2009markov} that the convergence time of such Markov chain can be bounded by the mixing time of a random walk on the same graph, which in turn has been proven~\cite{beveridge2013exact} to be at most quadratic on the number of nodes for paths. 
\miguel{However, the question of whether a static path is a worst case for Counting in some model of \ADN remains open.}

%\pichskip{15pt}% Horizontal gap between picture and text
%\parpic[l][t]{%
%  \begin{minipage}{.5\textwidth}
%  \large
%       \resizebox{\textwidth}{!}{\input{./plots75nodes/plot_pathByInterval}}
%    \caption{\IC time performance for a path topology.}
%       \label{fig:pathByInterval}%
%  \end{minipage}
%}

As a byproduct, our simulations provided insight on the impact of network dynamics in the dissemination of information by gossip-based protocols. Indeed, our results showed that on average network changes speed-up convergence. That is, as long as the effect is uniform throughout the network, highly dynamic topologies help rather than being a challenge as in a worst-case theoretical analysis.

%Open problems left for future work include further validating the conclusions in this work by simulating \IC using real traces, taken for instance from cellular networks, or by carrying out experiments in real networks. Also, once the network size is known, studying more complex problems in \ADNs is the natural next step. On the theoretical side, proving a polynomial upper bound on the time for counting remains an enticing open question.

\section{Conclusions and Open Problems}
\label{sec:conclude}

The problem of Counting in \ADNs is challenging because lack of identifiers and changing topology make difficult to decide if a new message has been received before from the same node. Counting protocols~\cite{opodisCounting,conscious,oracle} overcome this challenge checking candidate sizes iteratively until the correct size is obtained. However, the worst-case convergence time of the gossip-based process used in those protocols has been bounded only exponentially, yielding only exponential guarantees for \IC~\cite{opodisCounting}. The situation is even worse in other works where a huge overestimate (proved in~\cite{spirakis}) is initially used as the candidate size, yielding a double-exponential protocol~\cite{conscious}.
Other Counting protocols do not terminate~\cite{conscious}, 
or terminate but do not provide running-time guarantees~\cite{oracle},
or compute only an exponential upper bound on the network size~\cite{spirakis}. 
Other heuristic protocols do not guarantee that the computation is correct~\cite{experimentalConscious}.

The main goal of this work was to show that in practice \IC may compute the size of \ADNs in a polynomial number of rounds of communication, obtaining always the correct result. Additionally, the extensive simulations carried out provided very interesting observations about the impact of topology and dynamics in performance, as detailed in Section~\ref{sec:discussion}. \mig{To the best of our knowledge, this is the first experimental study of the practical time complexity of a Counting protocol that computes the exact count in \ADNs.}

Open problems left for future work include further validating the conclusions in this work by simulating \IC using real traces, taken for instance from cellular networks, or by carrying out experiments in real networks. Also, once the network size is known, studying more complex problems in \ADNs is the natural next step. On the theoretical side, proving a polynomial upper bound on the time for counting remains an enticing open question.

%\section*{Acknowledgments}
%\label{section:ack}
%We thank David Joiner for assisting us in using the Kean Terascale Cluster (KTC) for our simulations.

\bibliographystyle{splncs03}
\bibliography{./Comprehensive_2010}

%\newpage
%\appendix
%\section*{Appendix}
%\input{appendix}

\end{document}